\newtheorem{theorem}{Theorem}

\newtheorem{proposition}[theorem]{Proposition}

\documentclass{PoS}

\title{Alternative multiplications and non-associativity in physics}

\ShortTitle{Alternative multiplications}

\author{\speaker{Vladislav Kupriyanov}\\

        CMCC-Universidade Federal do ABC, Santo Andr\'e, SP, Brazil;\\
        
        Tomsk State University, Tomsk, Russia.\\

        E-mail: \email{vladislav.kupriyanov@gmail.com}}

\abstract{Some physical systems like the quantum mechanics with magnetic charges or field theoretical models appearing in the context of string theory are formulated in terms of non-associative algebras. Hence, demand non-associative star products for its description. However, unlike its associative counterpart the situation with non-associative star products is quite unclear. At the moment it is not evident which condition should be used instead of the associativity in the definition of the non-associative star multiplication. In mathematics, the natural generalization of the associativity is the requirement that the multiplication should be alternative, the associator of three elements should vanish if each two of them are equal. We show that for the alternative closed star product the integrated associator vanishes. In particular, it means that on-shell there will be no violation of associativity in string scattering amplitudes. We discuss the examples of the alternative and closed star products. It is constructed the star product corresponding to the algebra of octonions.}

\FullConference{Proceedings of the Corfu Summer Institute 2015 "School and Workshops on Elementary Particle Physics and Gravity"\\

                 1-27 September 2015\\

                 Corfu, Greece}

\begin{document}

\section{Introduction}\label{sec-intro}

Star product is an important mathematical tool in theoretical physics with inumerous applications. Originally it appeared in the context of deformation quantization approach to quantum mechanics (QM) \cite{BFFLS} and is defined as a formal deformation of the usual point-wise multiplication of smooth functions on some manifold $\mathcal{M}$,
\begin{equation}
f\star g = f \cdot g +\sum_{r=1}^\infty (i\alpha)^r C_r(f,g) \,,\label{star}
\end{equation}
where $\alpha$ is a deformation parameter and $C_r$ denote some bi-differential operators.  This product
provides a quantization of the bracket, $\{ f,g\}=P^{ij}(x) \partial_i f \ \partial_j g$, if
\begin{equation}
C_1(f,g)-C_1(g,f)= 2\{ f,g\}\,.\label{quant}
\end{equation}
In the classical setting, the associativity of star products, i.e., the requirement that the associator of any three functions should vanish,
\begin{equation}
A(f,g,h)=f\star (g\star h)-(f\star g)\star h=0  \,,\label{assoc}
\end{equation}
reflects the associativity of compositions of quantum mechanical operators. The condition (\ref{assoc}) implies the Jacobi identity (JI) on the bi-vector field $P^{ij}(x)$
and allows to proceed to higher orders, $C_r(f,g)$, $r>1$. A milestone result in deformation quantization is the Formality Theorem by Kontsevich \cite{Kontsevich} that demonstrates the existence of associative star products and states that each two star products $\star$ and $\star^\prime$ corresponding to the same  Poisson (i.e., satisfying JI) bi-vector $P^{ij}(x)$ are related by the gauge transformation
\begin{equation}
f\star^\prime g=\mathcal{D}^{-1}\left( \mathcal{D}f\star   \mathcal{D}g \right), \,\,\,\,\,\,\,\mathcal{D}=1+\mathcal{O}(\alpha)\,.\label{gauge}
\end{equation}
Gauge equivalence classes of star products are classified by Poisson structures on $\mathcal{M}$.

The interest to star products was boosted by the discovery \cite{ChuHo,Schom} that the coordinates
of string endpoints attached to a Dirichlet brane in a background $B$-field do not commute, and
thefore, star products are natural to describe correlation functions. It was found a bit later
\cite{CoSch,MK1,MK2}, that in a non-constant $B$-field, the product should be not only
noncommutative, but also nonassociative. More recently, similar effects were discovered in
the closed string sector. It was demonstrated \cite{Lust1,Lust2,Lust3,Andriot} that the presence of a
non-geometric $R$-flux leads to a twist of the Poisson structure and to nonasociativity of
corresponding star products. Note, that the mechanism of breaking the associativity is very
similar to the one that was found in the background of a magnetic monopole \cite{Jackiw}.
An extended discussion on the relations between the magnetic field and the nonassociativity may be found
in \cite{BaLu}.

Despite this renewed interest the overall situation with non-Poisson structures and nonassociative star products remains by far less clear than with their
associative counterparts. In this case, the condition (\ref{quant}) can still be imposed, but it is unclear which restriction should be used instead of the associativity (\ref{assoc}) to define the higher order terms in (\ref{star}). Why, for instance, one cannot simply set, $C_r(f,g)=0$, $r>1$, and define the star product as
\begin{equation}
f\star_0g=\ f \cdot g +i\alpha\{ f,g\} \,?\label{zero}
\end{equation}
To answer this question, it is reasonable to follow the standard logic of deformation quantization and to search for the physical restrictions on the nonassociative models. The requirement that the physical measurements should not feel the violation of the associativity may serve as such a restriction. In string theory it can be achieved if, on-shell, when the string equation of motion are satisfied, the integrated associator vanishes, see \cite{MK2, Blum}.

In the next section we show that for the {\it closed alternative} star product the integrated associator of three smooth functions, vanishing on the infinity, vanishes. The star product is called closed, if for Schwartz functions $f$ and $g$, we have
\begin{equation}\label{trace}
    \int f\star g=   \int  f\cdot g.
\end{equation}
The multiplication map is alternative if the associator vanishes whenever two of its arguments are equal. It means that the associator of any three elements $A(f,g,h)$ must be totally antisymmetric. Since, any associative algebra is automatically alternative, the later condition is a natural generalization of the associativity. In particular, the star product (\ref{zero}) is closed with appropriate integration measure, but it is not alternative. One may check that for this star multiplication the integrated associator will not vanish. So, it is not suitable for our aims.

We discuss the existence of the alternative star products corresponding to the arbitrary antisymmetric bi-vector $P^{ij}(x)$. Then we show that the requirement for the star product to be alternative is invariant under the gauge transformations (\ref{gauge}), just like it happens in the associative case. Different star products providing quantization of the same bracket reflect different quantization prescriptions of the same classical model. Physical consequences of each choice of the quantization scheme may be different, however the basic principles of the consistency must be the same. That is why, it is important that the condition restricting the higher order terms in (\ref{star}) should distinguish the whole class of admissible quantizations.

We consider the examples of alternative and non-associative star products. In the Sec. 3, it is proposed the explicit expression of the star product corresponding to the algebra of octonions. Then we provide the gauge transformation $\mathcal{D}$ relating this star product to the closed one. Finally, in the Sec. 4, we discuss the general case of bi-vector $P^{ij}(x)$. First we describe the construction of non-associative weakly-Hermitian (i.e., satisfying, $ (x^j\star f)^\ast=f^\ast \star x^j$, for all $x^j$) Weyl star product \cite{starpr} and show that it is weakly alternative, meaning that the associator $A(x^i,x^j,x^k)$ is totally antisymmetric. Our conjecture is that the Hermitian, $(f\star g)^\ast=g^\ast \star f^\ast,$ Weyl star product is alternative, $A(f,g,h)$ is antisymmetric for any functions $f$, $g$ and $h$. Then we derive the first orders of the gauge operator $\mathcal{D}$ mapping the proposed star product to the closed one. Thus, in lowest orders we give the explicit formulas for the closed alternative star product providing quantization of a (non)-Poisson bracket $\{ f,g\}=P^{ij}(x) \partial_i f \ \partial_j g$.

\section{Alternative star products}

{\bf Definition.} According to the above given definition, the star product is alternative if the associator of any three functions vanishes whenever two of them are equal, i.e.,
\begin{eqnarray}
A(f,f,g)=0, \\A(g,f,f)=0. \nonumber
\end{eqnarray}
These two identities together imply that the associator $A(f,g,h)$ must be totally antisymmetric. On the other hand, the Jacobiator,
\begin{equation}\label{jac}
  [f,g,h]_\star=[f,[g,h]_\star]_\star+[h,[f,g]_\star]_\star+[g,[h,f]_\star]_\star,
\end{equation}
where $[f,g]_\star$ stands for the star commutator is antisymmetric by definition. Also, the definitions (\ref{assoc}) and (\ref{jac}) mean that
\begin{equation}\label{jass}
    [f,g,h]_\star=A(f,g,h)-A(f,h,g)+A(h,f,g)-A(h,g,f)+A(g,h,f)-A(g,f,h).
\end{equation}
Since, for the alternative star product the associator should be totally antisymmetric, one finds
\begin{equation}\label{alt}
f\star (g\star h)-(f\star g)\star h=  \frac{1}{6}  [f,g,h]_\star,\,\,\,\,\,\forall f,g,h.
\end{equation}
That is, the star product is alternative {\it if and only if} the relation (\ref{alt}) holds.

{\bf Existence.} The first point we would like to address here is whenever the alternative star product exists for each non-Poisson bracket (\ref{quant}), or the equation (\ref{alt}) implies some restrictions on the bi-vector $P^{ij}(x)$. The eqs. (\ref{star}) and (\ref{quant}) give
\begin{equation}\label{PB}
 \{ f,g\}=      \left. \frac{f\star g-g \star f}{2i\alpha}\right|_{\alpha=0}.
\end{equation}
For the associative star products the equation (\ref{assoc}) already in the second order, $\mathcal{O}(\alpha^2)$, implies the JI,
\begin{equation}\label{JI}
 \{ f,\{g,h\}\}+ \{ h,\{f,g\}\}+ \{ g,\{h,f\}\}=0,
\end{equation}
which is the consistency condition in this case. It happens that no other restrictions appear and on any Poisson manifold there exists an associative star product \cite{Kontsevich}.

Applying the same logic to the alternative star multiplication and using the condition (\ref{alt}) instead of the associativity equation (\ref{assoc}) we end up with
\begin{equation}\label{JI}
 \{ f,\{g,h\}\}+ \{ h,\{f,g\}\}+ \{ g,\{h,f\}\}=\left.\frac{ [f,g,h]_\star}{-\alpha^2}\right|_{\alpha=0},
\end{equation}
which is nothing but the identity due to  (\ref{star}), (\ref{quant}) and (\ref{jac}). Apparently, in the lower orders the alternative star products do not imply any restriction on the antisymmetric bi-vector $P^{ij}(x)$. To prove the existence of the star product in all orders one may provide an explicit construction for it. The candidate of such a construction is the Weyl star product \cite{starpr} which we discuss in the last section.

{\bf Gauge invariance.} Suppose that the star products $\star$ and $\star^\prime$ are related by the gauge transformation (\ref{gauge}). Let us calculate
\begin{eqnarray}
A^\prime(f,g,h)&=&f\star^\prime (g\star^\prime h)-(f\star^\prime g)\star^\prime h \label{assoc1}\\
&=&\mathcal{D}^{-1}\left\{ \mathcal{D}f\star   \mathcal{D}\left[\mathcal{D}^{-1}\left( \mathcal{D}g\star   \mathcal{D}h \right)\right]\right\}-\mathcal{D}^{-1}\left\{ \mathcal{D}\left[\mathcal{D}^{-1}\left( \mathcal{D}f\star   \mathcal{D}g \right)\right]\star   \mathcal{D}h\right\}\,.\nonumber
\end{eqnarray}
Since, the gauge operator $\mathcal{D}$ is linear, one finds
\begin{equation}
A^\prime(f,g,h)=\mathcal{D}^{-1}A (\mathcal{D}f,\mathcal{D}g,\mathcal{D}h) \,.\label{assoc2}
\end{equation}
For the Jacobiators, calculated with respect to the $\star$ and $\star^\prime$, one finds the similar relation,
\begin{equation}
 [f,g,h]_{\star^\prime}=\mathcal{D}^{-1} [\mathcal{D}f,\mathcal{D}g,\mathcal{D}h]_\star \,.\label{jac2}
\end{equation}
The combination of (\ref{assoc2}) and (\ref{jac2}) gives
\begin{equation}\label{alt1}
   [f,g,h]_{\star^\prime}-6A^\prime(f,g,h)= \mathcal{D}^{-1}\left[ [\mathcal{D}f,\mathcal{D}g,\mathcal{D}h]_\star-6A(\mathcal{D}f,\mathcal{D}g,\mathcal{D}h)\right]=0,
\end{equation}
because of the equation (\ref{alt}) and the invertibility of the operator $\mathcal{D}=1+\mathcal{O}(\alpha)$. This means that the alternativity condition is invariant under the gauge transformations (\ref{gauge}). The situation is similar to the associative case.

{\bf Integration.} The definition of the consistent trace functional (integration) on the algebra of functions endowed with a star product is an important problem. Obviously, for the closed star products, $  \int f\star g=   \int  f\cdot g$, the integrated star commutator and the integrated Jacobiator vanish.
If the star product is also alternative, due to the definition (\ref{alt}),
\begin{equation}\label{trace1}
    \int (f\star g)\star h=   \int  f\star (g\star h).
\end{equation}
{\it For the alternative closed star product the integrated associator vanishes.} This result is of great importance in physics, see e.g. \cite{MK2, Blum}. In particular, it means that on-shell, i.e. if the string equation of motion are satisfied, there will be no violation of associativity in string scattering amplitudes.

{\bf Examples.} As it was already mentioned in the introduction, any associative star product is automatically alternative, it can be considered as a trivial example. 

The following algebra of brackets appears in the context of both flat non-geometric $R$-flux in string theory and QM with magnetic charges,
\begin{equation}\label{rb}
\{x^i, x^j\}=R^{ijk}p_k,\,\,\,\,\{x^i,p_j\}=\delta^i_j,\,\,\,\,\{p_i,p_j\}=0,
\end{equation}
where $R^{ijk}$ is totally antisymmetric and $i,j,k=1,2,3$. In \cite{MSS1} it was demonstrated that the corresponding to (\ref{rb}) star product,
\begin{equation}\label{starp}
f\star_pg=\mu_2\left(\exp\left[\frac{i\alpha}{2}R^{ijk}p_k\partial_i\otimes\partial_j\right]\exp\left[\frac{i\alpha}{2}(\partial_i\otimes\tilde\partial^i-\tilde\partial^i\otimes\partial_i)\right]
f\otimes g\right),
\end{equation}
where $\mu_2$ is a point-wise multiplication map, and $\tilde\partial^i=\partial/\partial p_i$, satisfy the requirement (\ref{alt}), i.e., it is alternative. Then in \cite{MSS2} it was shown that (\ref{starp}) is closed and satisfy (\ref{trace1}). In the terminology of \cite{MSS2} the properties (\ref{trace}) and (\ref{trace1}) are refered to as $2$-cyclicity and $3$-cyclicity correspondingly.

\section{Star product for octonions}

Probably, the most known example of the alternative, but non-associative algebra are the octonions. Every octonion $\mathcal{X}$ can be written in the form
\begin{equation}\label{oct}
\mathcal{X}=\mathcal{X}_0e_0+\mathcal{X}_1e_1+\mathcal{X}_2e_2+\mathcal{X}_3e_3+\mathcal{X}_4e_4+\mathcal{X}_5e_5+\mathcal{X}_6e_6+\mathcal{X}_7e_7,
\end{equation}
where $\mathcal{X}_i$ are the real coefficients, $e_0$ is the scalar element, and the imaginary unit octonions $e_i$ satisfy the following multiplication rule:
\begin{equation}\label{oct1}
e_ie_j=-\delta_{ij}e_0+\varepsilon_{ijk}e_k,
\end{equation}
with $\varepsilon_{ijk}$ being a completely antisymmetric tensor of the rank three with a positive value $+1$ when $ijk = 123,$ $145,$ $176,$ $246,$ $257,$ $347,$ $365$, and $i,j,k=1\dots7$. The octonions are neither commutative, nor associative. The commutator algebra of the octonions is given by
\begin{equation}\label{oct2}
[e_i,e_j]:=e_ie_j-e_je_i=2\varepsilon_{ijk}e_k.
\end{equation}
In seven dimensions one has
\begin{equation}\label{epsilon7}
\varepsilon_{ijk}\varepsilon_{lmk}=\delta_{il}\delta_{jm}-\delta_{im}\delta_{jl}+\varepsilon_{ijlm},
\end{equation}
where $\varepsilon_{ijlm}$ is a completely antisymmetric tensor of the rank four with a positive value $+1$ when $ijlm = 1247,$ $1265,$ $1436,$ etc., One may also represent the tensor with four indices $\varepsilon_{ijlm}$ as a dual to the tensor with three indices $\varepsilon_{kpr}$ through 
\begin{equation}\label{epsilon8}
\varepsilon_{ijlm}=\frac{1}{6}\varepsilon_{ijlmkpr}\varepsilon_{kpr},
\end{equation}
where $\varepsilon_{ijlmkpr}$ is the Levi-Civita symbol in seven dimensions, normalized as $\varepsilon_{1234567}=+1$, see e.g. \cite{Toppan} for details. Taking into account (\ref{epsilon7}), for the Jacobiator we get
\begin{equation}\label{oct3}
[e_i,[e_j,e_k]]+cycl(ijk)=6\varepsilon_{ijkl}e_l.
\end{equation}
In this section we are interested in the quantization of the bracket:
\begin{equation}\label{oct4}
\{x^i,x^j\}=\varepsilon^{ijk}x^k, \,\,\,\,\,i,j,k=1\dots7,
\end{equation}
representing the algebra of imaginary octonions.

{\bf Vector star product.} In \cite{FL} in three dimensional case it was proposed the following multiplication map. For each two vectors $\vec p_1,\vec p_2$ from the unit ball, $p_ip^i\leq1$, it is assigned the vector $\vec p_1\circledast\vec p_2$ by the rule,
\begin{equation}\label{vstar}
\vec p_1\circledast\vec p_2={\sqrt{1-| \vec p_1|^2} }\,\,\vec p_2+\sqrt{1-| \vec p_2|^2}\,\,\vec p_1-\vec p_1\times\vec p_2,
\end{equation}
    where $| \vec p|=\sqrt{ \vec p\cdot \vec p}$ is the Euclidean vector norm, $\vec p_1\cdot \vec p_2$  and $\vec p_1\times\vec p_2$ stand for the dot and the cross products correspondingly. This multiplication is noncommutative, but as we will see below is associative. Since,
    \begin{equation}
1-|\vec p_1\circledast\vec p_2|^2=\left(\sqrt{(1-|\vec  p_1|^2)(1-|\vec p_2|^2)} -\vec p_1\cdot\vec p_2\right)^2\geq0,
\end{equation}
the resulting vector $\vec p_1\circledast\vec p_2$ also belongs to the unit ball. The above multiplication also can be defined in seven dimensions, however in this case it will be non-associative, but alternative.

For the associator of three vectors one finds
\begin{eqnarray}\label{assvstar}
A(\vec p_1,\vec p_2,\vec p_3)&=&\left(\vec p_1\circledast\vec p_2\right)\circledast\vec p_3-\vec p_1\circledast\left(\vec p_2\circledast\vec p_3\right)\\
&=&\left(\vec p_1\times\vec p_2\right)\times\vec p_3-\left(\vec p_1\cdot\vec p_2\right)\vec p_3-\vec p_1\times\left(\vec p_2\times\vec p_3\right)+\left(\vec p_2\cdot\vec p_3\right)\vec p_1.\nonumber
\end{eqnarray}
In three dimensions due to the identity, $\vec p_1\times\left(\vec p_2\times\vec p_3\right)=\left(\vec p_1\cdot\vec p_3\right)\vec p_2-\left(\vec p_1\cdot\vec p_2\right)\vec p_3,$ the above expression vanishes, which means that the vector star product (\ref{vstar}) is associative. In seven dimensions the coordinates of the vector $\vec p_1\times\vec p_2$ can be written as
\begin{equation}
(\vec p_1\times\vec p_2)_i=\varepsilon_{ijk}p_1^jp_2^k,
\end{equation}
where $\varepsilon_{ijk}$ is defined after the eq. (\ref{oct1}). Taking into account (\ref{epsilon7}) one obtains for the associator (\ref{assvstar}) written in the components,
\begin{equation}
A(\vec p_1,\vec p_2,\vec p_3)_i=-2\varepsilon_{ijlm}p_1^jp_2^lp_3^m.
\end{equation}
It is totally antisymmetric, meaning that in seven dimensions the vector star product (\ref{vstar}) is no longer associative, but alternative.

To generalize the vector star product (\ref{vstar}) for the whole Euclidean space $V$ we introduce the map
\begin{equation}
 \vec p=\frac{\sin\left(\frac{\alpha}{2}|\vec k|\right)}{|\vec k|}\vec k,\,\,\,\,k^i\in\mathbb{R}.
\end{equation}
The inverse map is given by
\begin{equation}
 \vec k=\frac{2\arcsin\left|\vec p\right|}{\alpha\left|\vec p\right|}\vec p.
\end{equation}
For each two vectors $\vec k_1,\vec k_2\in V$ we set:
\begin{equation}
\label{Bk}
\mathcal{ B}(\vec k_1,\vec k_2)=\left.\frac{2\arcsin\left|\vec p_1\circledast\vec p_2\right|}{\alpha\left|\vec p_1\circledast\vec p_2\right|}\vec p_1\circledast\vec p_2\right|_{ \vec p_a=\vec k_a\sin\left(\frac{\alpha}{2}|k_a|\right)/|k_a|},\,\,\,a=1,2.
\end{equation}
From the definition (\ref{Bk}) one immediately finds the following properties:
\begin{description}
\item[i)]  $\mathcal{ B}(\vec k_1,\vec k_2)=-\mathcal{ B}(-\vec k_2,-\vec k_1)$;
\item[ii)] $\mathcal{ B}(\vec k,0)=\mathcal{ B}(0,\vec k)=\vec k$;
\item[iii)]  $\mathcal{ B}(\vec k_1,\vec k_2)=\vec k_1+\vec k_2-{\alpha}\vec k_1\times\vec k_2+\dots$;
\item[iv)] in three dimensions the vector multiplication (\ref{Bk}) is associative, $$\mathcal{ B}(\mathcal{ B}(\vec k_1,\vec k_2),\vec k_3)=\mathcal{ B}(\vec k_1,\mathcal{ B}(\vec k_2,\vec k_3));$$ while, in seven dimensions it is not associative, but alternative.
\end{description}
The derivation of the expression for $\mathcal{ B}(\vec k_1,\vec k_2)$ in three space dimensions and its relation with BCH formula for $SU(2)$ can be found in \cite{FM,OR13}.

{\bf Star product.} Let us define the star product as
\begin{equation}\label{w1}
f\star g(x) =\int \frac{d^{7}k_1}{\left( 2\pi
\right) ^{7}} \frac{d^{7}k_2}{\left( 2\pi
\right) ^{7}}\tilde{f}\left( k_1\right)\tilde{g}\left( k_2\right)e^{i\mathcal{ B}(\vec k_1,\vec k_2)\cdot{\vec x}},
\end{equation}
where $\tilde{f}$ stands for the Fourier transform of the function $f$. Due to the properties of the vector multiplication $\mathcal{ B}(\vec k_1,\vec k_2)$, the introduced star product is hermitian,
\begin{equation}\label{41}
    (f\star g)^\ast=g^\ast \star f^\ast,
\end{equation}
satisfies the stability of the unity, i.e., $f\star 1=1\star f=f(x)$. It provides the quantization of the bracket (\ref{oct4}),
\begin{equation}\label{41}
       \left. \frac{f\star g-g \star f}{2i\alpha}\right|_{\alpha=0}=x^i\varepsilon_{ijk}\partial_jf\partial_kg.
\end{equation}
The property ($\bf{iv}$) implies that the star product (\ref{w1}) is alternative. Consequently, the star commutator $[f,g]_\star$ satisfy the Malcev identity.

Let us calculate $x^i\star f$ using (\ref{w1}). Since the Fourier transform of $x^i$ is the derivative of a Dirac delta function, we get
\begin{eqnarray}
x^i\star f &=&\int \frac{d^{7}k_1}{\left( 2\pi
\right) ^{7}} \frac{d^{7}k_2}{\left( 2\pi
\right) ^{7}}(2\pi i )^7\left(\partial^i_{k_1}\delta(k_1)\right)\tilde{f}\left( k_2\right)e^{i\mathcal{ B}(\vec k_1,\vec k_2)\cdot{\vec x}}\\
&=&-\int \frac{d^{7}k_1}{\left( 2\pi
\right) ^{7}} d^{7}k_2 x^l\frac{ \partial  \mathcal{ B}_l(\vec k_1,\vec k_2)}{\partial k_1^i}\delta(k_1)\tilde{f}\left( k_2\right)e^{i\mathcal{ B}(\vec k_1,\vec k_2)\cdot{\vec x}}\nonumber\\
&=&-\int \frac{d^{7}k_2}{\left( 2\pi
\right) ^{7}}  x^l\left.\frac{ \partial  \mathcal{ B}_l(\vec k_1,\vec k_2)}{\partial k_1^i}\right|_{k_1=0}\tilde{f}\left( k_2\right)e^{i\mathcal{ B}(0,\vec k_2)\cdot{\vec x}}\nonumber
\end{eqnarray}
After some algebra one may see that
\begin{eqnarray*}
&&\left.\frac{ \partial \mathcal{ B}_l(\vec k_1,\vec k_2)}{\partial k_1^i}\right|_{k_1=0}=\\
&&-\alpha \varepsilon^{ilm}k_2^m+\delta^{il}\frac{\alpha}{2}|k_2|\cot\left(\frac{\alpha}{2}|k_2|\right)
+\frac{k_2^ik_2^l}{|k_2|^2}\left(\frac{\alpha}{2}|k_2|\cot\left(\frac{\alpha}{2}|k_2|\right)-1\right).\nonumber
\end{eqnarray*}
Then, taking into account ({\bf ii}) and integrating over $k_2$ we conclude that
\begin{eqnarray}
x^i\star f &=&\left\{x^i+\frac{i\alpha}{2}\varepsilon^{ijk}x^k\partial_j\right.\\
&+&\left.
  (x^i\Delta-x^l\partial_l\partial_i)\Delta^{-1}\left[\frac{\alpha}{2}\sqrt{\Delta}\coth\left(\frac{\alpha}{2}\sqrt{\Delta}\right)-1\right]\right\}\triangleright f,\nonumber \label{poly}
\end{eqnarray}
where $\Delta=\partial_i\partial^i$.

For the Jacobiator one finds,
\begin{equation}\label{oct5}
[x^i,x^j,x^k]_\star=-3\alpha^2\varepsilon^{ijkl}x^l,
\end{equation}
which is in agreement with (\ref{oct3}).

{\bf Integration.} To introduce the trace functional (integration) on the algebra of the star product first we note that $  \partial_{i}\left( \varepsilon^{ijk}x^{k}  \right)  =0$, due to the antisymmetry of $\varepsilon^{ijk}$. Thus, for functions $f$ and $g$ vanishing on the infinity and the bracket (\ref{oct4}) one has, $\int  d^{7}x \{f,g\}=0$. That is, it is reasonable to take the canonical integration as the corresponding trace functional. However, the star product (\ref{w1}) is not closed,
 \begin{equation}\label{oct6}
 \int  d^{7}x [f\star g- f\cdot g]\neq0 .\end{equation}
To overcome this difficulty one may search for the gauge equivalent, in the sense (\ref{gauge}), star product which would be closed with respect to the introduced integration. In \cite{KV15} we did it for the star product corresponding to the algebra $\mathfrak{su}(2)$. The procedure of finding the gauge operator for the star product (\ref{w1}) is absolutely analogous to \cite{KV15}, except for some coefficients because of the different dimensions. Here we give only the final result. The star product
\begin{equation}\label{oct7}
f\circ g= \mathcal{D}^{-1}\left( \mathcal{D}f\star   \mathcal{D}g \right),
\end{equation}
where
\begin{equation}\label{oct8}
\mathcal{D}=\left( \frac{2\sinh\left(\frac{1}{2}\alpha\sqrt{\Delta}\right)}{\alpha\sqrt{\Delta}}\right)^{\frac{1}{3}},
\end{equation}
is closed with respect to the integration,
 \begin{equation}\label{oct9}
 \int  d^{7}x f\circ g= \int  d^{7}x f\cdot g .
 \end{equation}
 By the construction, (\ref{oct7}) is alternative. Consequently, for this star product the integrated associator vanishes.

\section{Weyl star product}

In this section we discuss the quantization of the arbitrary (non)-Poisson bracket, following \cite{starpr}. To start with let us introduce some definitions. 
\begin{itemize}
\item The $\star$ is called {\it Weyl star product}, if it satisfies
  \begin{equation}\label{weyl}
    (x^{i_1}\dots x^{i_n})\star f=\sum_{P_n} \frac 1{n!} P_n( x^{i_1}\star(\dots \star (x^{i_n}\star f)\dots)\,.
\end{equation}
where $P_n$ denotes a permutation of $n$ elements. E.g., for the product of $x^ix^j$,
 \begin{equation}\label{weyl1}
  (x^ix^j)\star f=\frac{1}{2}\left(x^i\star(x^j\star f)+x^j\star(x^i\star f)\right).
  \end{equation}
\item {\it Weak Hermiticity} means that for all $x^j$,   \begin{equation}\label{wher}
    (x^j\star f)^\ast=f^\ast \star x^j.
\end{equation}
\item The star product is {\it strictly triangular}, if the r.h.s. of
\begin{equation}\label{ST}
\frac{x^i\star x^j-x^j \star x^i}{2i\alpha}=P^{ij}(x),
\end{equation}
does not depend on $\alpha$.
\end{itemize}
\begin{proposition}
For any bivector field $P^{ij}$ there is unique weakly Hermitian strictly triangular Weyl star product satisfying the stability of
unity condition, $f(x)\star 1=1\star f(x)=f(x)$.
\end{proposition}
The proof of this statement is constructive, see \cite{starpr}. We present recursion relations that allow to compute this star product to any given order. In particular, up to the third order we have:
\begin{eqnarray}
&&(f\star g)(x)=f\cdot g+i\alpha P ^{ij}\partial
_{i}f\partial _{j}g  \label{starw} \\
&&-\frac{\alpha ^{2}}{2}P^{ij}P ^{kl}\partial _{i}\partial
_{k}f\partial _{j}\partial _{l}g-\frac{\alpha ^{2}}{3}P ^{ij}\partial
_{j}P ^{kl}\left( \partial _{i}\partial _{k}f\partial _{l}g-\partial
_{k}f\partial _{i}\partial _{l}g\right) \nonumber \\
&&-i\alpha^3 \left[
\frac{1}{3}P ^{nl}\partial _{l}P ^{mk}\partial _{n}\partial
_{m}P ^{ij}\left( \partial _{i}f\partial _{j}\partial _{k}g-\partial
_{i}g\partial _{j}\partial _{k}f\right) \right.  \nonumber \\
&&+ \frac{1}{6}  P^{nk}\partial_nP^{jm}\partial_mP^{il}
\left(\partial_i\partial_jf\partial_k\partial_lg-\partial_i\partial_jg\partial_k\partial_lf\right)   \nonumber \\
&&+\frac{1}{3}P ^{ln}\partial _{l}P ^{jm}P ^{ik}\left(
\partial _{i}\partial _{j}f\partial _{k}\partial _{n}\partial _{m}g-\partial
_{i}\partial _{j}g\partial _{k}\partial _{n}\partial _{m}f\right)
\nonumber \\
&&+\frac{1}{6}P ^{jl}P ^{im}P ^{kn}\partial _{i}\partial
_{j}\partial _{k}f\partial _{l}\partial _{n}\partial _{m}g  \nonumber \\
&&\left. +\frac{1}{6}P ^{nk}P ^{ml}\partial _{n}\partial _{m}P
^{ij}\left( \partial _{i}f\partial _{j}\partial _{k}\partial _{l}g-\partial
_{i}g\partial _{j}\partial _{k}\partial _{l}f\right)\right] +\mathcal{O}\left( \alpha ^{4}\right)~.  \nonumber
\end{eqnarray}
Up to this order the star product is hermitian and satisfies (\ref{alt}) for any three function $f,g$ and $h$, i.e., alternative.

Let us discuss an important property of the introduced star product which holds true in all orders.
For the Weyl star product one has by (\ref{weyl})
\begin{equation}\label{xxx}
  (x^ix^j)\star x^k=\frac{1}{2}\left(x^i\star(x^j\star x^k)+x^j\star(x^i\star x^k)\right).
\end{equation}
On the other hand the stability of the unity,
\begin{equation}
  \frac{1}{2}\left(x^i\star x^j+x^j\star x^i\right)=x^ix^j,
\end{equation}
implies
\begin{equation}\label{j1}
 \left(x^i\star x^j\right)\star x^k+\left(x^j\star x^i\right)\star x^k=  x^i\star(x^j\star x^k)+x^j\star(x^i\star x^k).
\end{equation}
Which means that
\begin{equation}\label{j2}
  A(x^i,x^j,x^k)+ A(x^j, x^i,x^k)=0,
\end{equation}
that is, the associator of coordinate functions $x^i$, $x^j$ and $x^k$ is antisymmetric in first two arguments. If the Weyl star product is weak-Hermitian, then considering complex conjugate of the equation (\ref{j1}) and 
using (\ref{wher}) we obtain
\begin{equation}\label{j3}
  x^k\star(x^j\star x^i)+x^k\star(x^i\star x^j)= \left(x^k\star x^j\right)\star x^i+\left(x^k\star x^i\right)\star x^j.
\end{equation}
That is, the associator of coordinate functions $x^k$, $x^j$ and $x^i$ is antisymmetric in the last two arguments,
\begin{equation}\label{j4}
  A(x^k,x^j,x^i)+ A(x^k, x^i,x^j)=0.
\end{equation}
Using now (\ref{j2}) and (\ref{j4}) in (\ref{jass}) we find
\begin{equation}\label{j5}
 x^i\star(x^j\star x^k)-(x^i\star x^j)\star x^k= \frac{1}{6}[x^i,x^j,x^k]_\star
\end{equation}
Let us call the star product satisfying (\ref{j5}) {\it weakly alternative}. We have proved that the weakly Hermitian Wey star product is weakly alternative.

The same way that the weak Hermiticity (\ref{wher}) does not necessarily implies 
Hermiticity,  $(f\star g)^\ast=g^\ast \star f^\ast$, the weakly alternative star product should not necessarily be alternative.  To make the Weyl star product hermitian one needs to relax the condition (\ref{ST}) 
and admit the corrections in $\alpha$ (renormalization) of the given bi-vector $P^{ij}(x)$,  $P^{ij}\rightarrow P_r^{ij}=P^{ij}+\alpha^2P^{ij}_2+\mathcal{O}\left( \alpha ^{4}\right)$. Our conjecture is that {\it hermitian Weyl star product is alternative.} 

For linear bi-vectors, $P^{ij}(x)=C^{ij}_kx^k$, the weakly hermitian star product is also hermitian and alternative in all orders. The explicit example of non-associative Weyl star product (\ref{w1}) which is hermitian and alternative was given in the previous section. 

{\bf Integration.} To define the integration measure $\mu(x)\neq0$, we impose the condition
\begin{equation}\label{measure}
 \partial_i\left(\mu\cdot P^{ij}\right)=0 .\end{equation}
 In this case, the integrated bracket of two Schwartz functions vanishes, $\int \{f,g\}=0$, implying that 
 \begin{equation}\label{m2}
 \int  d^{N}x\mu [f\star g- f\cdot g]=\mathcal{O}\left( \alpha ^{2}\right) .\end{equation}
In open string theory one may choose, $\mu=\sqrt{\det(g+\mathcal{F})}$, as a Born-Infeld measure.In this case, (\ref{measure}) are the equations of motion for the corresponding gauge potencial  \cite{MK2, Blum}. The problem is that for the Weyl star product (\ref{starw}), already in the second order, $\mathcal{O}\left( \alpha ^{2}\right)$, the l.h.s. of (\ref{m2}) does not vanish. To give the precise expression of (\ref{m2}) we note that because of the identities,
\begin{equation}\int d^{N}x\partial _{i}f\partial _{l}\left( \mu
\Pi ^{ilk}\right) \partial _{k}g=0,\,\,\,\,\partial _{l}\left( \mu
P ^{lj}\partial _{j}P ^{ki}\right) =0,\end{equation}
one has,
\begin{equation}\label{m3}
\int d^{N}x\partial _{i}f\partial _{l}\left( \mu
P ^{ij}\partial _{j}P ^{lk}\right) \partial _{k}g=\int d^{N}x\partial _{i}g\partial _{l}\left( \mu
P ^{ij}\partial _{j}P ^{lk}\right) \partial _{k}f,
\end{equation}
i.e., the matrix 
\begin{equation}\label{m4}
b^{ik}(x)=\partial _{i}f\partial _{l}\left( \mu P ^{ij}\partial _{j}P ^{lk}\right) \partial _{k}g
\end{equation} 
is symmetric in the indices $i,k$ up to the surface terms. Now we find,
\begin{equation}\label{m5}
 \int  d^{N}x\mu [f\star g- f\cdot g] =-\frac{\alpha ^{2}}{6}\int d^{N}x\partial _{i}fb^{ik} \partial _{k}g+\mathcal{O}\left( \alpha ^{3}\right) .
\end{equation}
To fix it, we will follow \cite{Kup15} and look for the gauge transformation $\mathcal{D}:\star\rightarrow \circ$, mapping the given star product $\star$ to the closed one $\circ$. Introducing the gauge operator 
\begin{equation}
\mathcal{D}=1-\frac{\alpha ^{2}}{12\mu }b^{ik}\partial _{i}\partial
_{k}+\mathcal{O}\left( \alpha ^{3}\right),
\end{equation}
one obtains the star product
\begin{eqnarray}\label{circ}
f\circ g&=& \mathcal{D}^{-1}\left( \mathcal{D}f\star   \mathcal{D}g \right)\\
&=& f\star  g +\frac{\alpha ^{2}}{12\mu}\partial _{l}\left( \mu P ^{ij}\partial
_{j}P ^{lk}\right)\left[\partial _{i}f\partial
_{k}g+\partial _{i}g\partial_{k}f\right]+\mathcal{O}\left( \alpha ^{3}\right).\nonumber
\end{eqnarray}
which is closed up to the third order in $\alpha$:
\begin{equation}
 \int  d^{N}x\mu [f\circ g- f\cdot g]=\mathcal{O}\left( \alpha ^{3}\right) ~.
\end{equation}
Since, the new star product (\ref{circ}) is gauge equivalent to the alternative one (\ref{starw}), it is also alternative. Hence, for this star product holds,
\begin{equation}\label{trace2}
    \int (f\circ g)\circ h=   \int  f\circ (g\circ h),
\end{equation}
the integrated associator vanishes. 

\section{Conclusions}

We argue that the violation of the associativity in the formulation of some physical models should not manifest itself on the physical observables. This in turn can be achieved if to require that the integrated associator vanishes. We show that alternative closed star products satisfy this condition and consequently are suitable to work with non-associative algebras appearing in physical context. Some more arguments in favor of alternative star products can be found in \cite{Bojowald1,Bojowald2}. In general case, discussing the quantization of brackets $\{f,g\}=P^{ij}(x)\partial_i f\partial_jg,$ we provide the procedure of the construction and the explicit expressions in first orders of the closed alternative star products. For the linear bracket corresponding to the algebra of imaginary octonions, we obtain the explicit formula for the star product in all orders.

Finally we note that there are some different approaches in treating the non-associative systems within the framework of deformation quantization. For exemple, one may require the associativity of the star product only for the physical observables \cite{LSh}, or introduce new elements to the original algebra which will make it associative \cite{Ho}.

The extended version of this contribution to the proceedings to the CORFU2015 containing more technical details and proofs is in preparation.

\acknowledgments

I appreciate the fruitful discussions with Sasha Pinzul, Francesco Toppan, Richard Szabo and Dieter L\"ust during the meeting in Corfu and scientific visits to Heriot-Watt University and MPI-Munich after the conference. I am also grateful Dima Vassilevich for helpful talks and remarks. This work was
supported by FAPESP and CNPq.

\end{document}